# Service-Oriented Simulation Framework: An Overview and Unifying Methodology

Wenguang Wang[+], Weiping Wang, Yifan Zhu and Qun Li

Department of Systems Engineering, College of Information Systems and Management, National University of Defense Technology Changsha 410073, China.

+Corresponding author: Wenguang Wang    Email: wgwangnudt@gmail.com

**Abstract:** The prevailing net-centric environment demands and enables modeling and simulation to combine efforts from numerous disciplines. Software techniques and methodology, in particular service-oriented architecture, provide such an opportunity. Service-oriented simulation has been an emerging paradigm following on from object- and process-oriented methods. However, the ad-hoc frameworks proposed so far generally focus on specific domains or systems and each has its pros and cons. They are capable of addressing different issues within service-oriented simulation from different viewpoints. It is increasingly important to describe and evaluate the progress of numerous frameworks. In this paper, we propose a novel three-dimensional reference model for a service-oriented simulation paradigm. The model can be used as a guideline or an analytic means to find the potential and possible future directions of the current simulation frameworks. In particular, the model inspects the crossover between the disciplines of modeling and simulation, service-orientation, and software/systems engineering. Based on the model, we present a comprehensive survey on several classical service-oriented simulation frameworks, including formalism-based, model-driven, interoperability protocol based, eXtensible Modeling and Simulation Framework (XMSF), and Open Grid Services Architecture (OGSA) based frameworks etc. The comparison of these frameworks is also performed. Finally the significance both in academia and practice are presented and future directions are pointed out.

## 1. Introduction

With the widespread application of modeling and simulation (M&S) techniques in military, education, aeronautics, astronautics, commerce, communication, manufacture, and other communities, many discipline-specific M&S frameworks have been built up. However, some deficiencies such as interoperability, composability, extensibility, agility, and reusability of current simulation frameworks have been revealed during the past decade. These frameworks do not adapt to the prevailing net-centric environment with the properties of distribution, collaboration, and sharing more and more. Therefore, there is an emerging need to combine efforts from numerous domains. This is driven by the extension of the application scope, the emergence of new technologies, and the need of net-centric simulation in Global Information Grid (GIG) [1].

Meanwhile, from the requirements' perspective, the enterprise application integration community needs the service-orientation idea and techniques to improve the reusability and agility of services and business processes. In the defense community, future wars are net-centric and full of uncertainty (Such as anti-terrorism). To deal with real time, uncertain decision, and application integration problems in a highly dynamic and agile sphere, it would be better and agile to compose and integrate "capability units" (services) than develop systems from scratch to response to quick tempo. Therefore, the Department of Defense (DoD) take service-oriented approaches as an enabler to improve the sharing of information, resources, and abilities, thereby increasing operational effectiveness. Consequently DoD proposes net-centric services strategy [2]. From the technique's perspective, Service-Oriented Architecture (SOA) [3-4] was proposed as a service-oriented framework to promote the reus-ability and interoperability of heterogeneous systems based on various operating systems, development platforms, programming languages and middlewares. Service-oriented paradigm is immerging as a new pattern following process-oriented and object-oriented ones in systems analysis and software development. Some new terms are springing up such as service-oriented science [5], computing [6], modeling [7], simulation [8-9], system engineering [10], software engineering [11-12], and high level architecture (HLA) [13].

Service-oriented paradigm brings new challenges to classical M&S frameworks towards net-centric environment. The use of SOA to extend the capability of M&S framework has attracted increasing attention [14]. In terms of simulation, various service-oriented simulation frameworks have been proposed or implemented by different institutes using different formalisms or techniques. These include formalism-based [15-16], model-driven [17], interoperability protocol based [13], Extensible Modeling and Simulation Framework (XMSF) [18] and Open Grid Services Architecture (OGSA) based [19] frameworks. However, the frameworks proposed so far generally focus on specific domains or systems and each has its pros and cons. They are capable of addressing different issues within service-oriented simulation from different viewpoints. It is increasingly important to develop a high level reference model that can describe and evaluate the progress of numerous service-oriented simulation frameworks. It is also important to identify the potential and possible future directions of current frameworks and facilitate research, development, improvements and the application of the old and new frameworks. Finally, it would lead to new solutions to the reusability, composability and interoperability of heterogeneous simulation resources.

In this work, we make significant extension compared with our preliminary research in a science letter [20]. We give detailed motivation and related concepts first. Then we propose the improved and detailed three-dimensional reference model. Taking this model as a taxonomy, we make a comprehensive survey on classical service-oriented simulation frameworks. We also perform detailed comparison from the viewpoint of one, two, and three dimensions (1D, 2D, and 3D) demanded by the 3D model. Finally, the research values are pointed out and future directions are recommended.

## 2. Concept Exploration

As the basis for later investigation, we first explore related concepts of service-oriented simulation.

### 2.1 Services

**Services** have different implications in different contexts. Dick et al. [21] and Savas et al. [22] summarized the definitions of services and their properties from the process, interaction, capability and operation etc. point of view. There are two prevailing definitions given by the World Wide Web Consortium (W3C) and the DoD. Within the domain of IT, the W3C define a service as *"an abstract resource that represents a capability of performing tasks that form a coherent functionality from the point of view of providers entities and requesters entities."* [23]. On the other hand, in defense community, the DoD define a service as *"a mechanism to enable access to one or more capabilities, where the access is provided using a prescribed interface and is exercised consistent with constraints and policies as specified by the service description."* [2]. Based on the definitions of a 'service' given above and by others, a lot of attention is paid to capability, utility, interface, and functionality aspects whereas implementation details are generally hidden. In this paper, we adopt the DoD's definition.

Given these characteristics, services have different taxonomies according to the types of capability, carrier, presentation, application scope and context etc. For example, the US net-centric services strategy classifies services as Core Enterprise Services (CES) and Communities of Interest (COIs) services [2]. Tolk et al. divide the services that accessing Common Reference Model (CRM) into atomic, composite, aggregate, and data mediation services from the perspective of model-based data engineering [24]. Suzić et al. propose the services taxonomy of Operational, System Management, Messaging, Registration and Discovery, Mediation, Collaboration, Information Assurance and Security, Storage, and Application Services in their core technical framework [25].

### 2.2 Simulation Services

Similarly, as a special kind of services, simulation services have different implications in different contexts. For example, in the context of High Level Architecture (HLA), simulation services may refer to the runtime infrastructure (RTI) services for models such as the time management, object management.

Taking Web as an implementation platform, Zhang et al. define **simulation services** as "Simulation Services are simulation components encapsulated with certain simulation applications or model logics, which have certain functions and are embodied as state-persistent Web services. The information and semantics of simulation services are described by Web service standards. The communication and interoperation among services are enabled by standard Web ser-vice protocols. Simulation services help to satisfy user's requirements through cooperation of all involved services." [26-27].

In this paper, we classify the general M&S capabilities into the definition of services. Therefore, we defined a **general simulation service** from the capability perspective as follows:

A 'simulation service' refers to the capability of M&S activities owned or implemented by abstract (i.e., conceptual) or concrete (i.e., implementation related) elements that can be used by other services.

Meanwhile, we regard Zhang's implementation-related definition [26-27] as the narrow sense. The definitions of simulation services in general and narrow senses are to facilitate the service-oriented concept, analysis, design and implementation.

### 2.3 Service-Oriented Simulation

The service-oriented simulation concept was originally proposed by Gustavsson et al. [8-9]. However, this concept is proposed from the viewpoints of Swedish Armed Force Enterprise Architecture Services, simulation, and software engineering. It has not evolved into the service-oriented simulation concept in general sense as a successor to object-oriented simulation [28]. Referring to the object-oriented, process-oriented, and event-oriented simulation concepts laid out by DoD [29], we define **service-oriented simulation** as:

*A simulation using a service-oriented paradigm in which the service and its capability are considered more important than the object, process or outcome etc. Service-oriented simulation focuses on the description, publish, composition in the lifecycle of services or simulation services. For example, in service-oriented war game simulation, an observation service pays more attention to observation capability than concrete processes or objects (human vision, telescope, radar etc.).*

There are three distinct, yet related concepts about service-oriented simulation. The first is "Service-based simulation", which means only using the basic SOA language/platform independent concept/properties (Corresponding to core issues, e.g., service provider and requestor) while not using the full potential of SOA (indirect addressing, broker, composition etc.). The second is "Service-oriented simulation", which uses the full potential of SOA especially broker and composition. (Corresponding to core and supporting issues). The third is "Service-oriented simulation engineering", which emphasize the use of engineering principles or approaches to the service-oriented simulation concept. (Corresponding to the general service-oriented simulation). In this work, we only use the general "service-oriented simulation" concept to represent the above three detailed definitions.

Service-oriented simulation has two research directions [14]. One is the application of M&S to SOA, e.g., using M&S techniques to address the analysis, design, evaluation, and testing problems in service-oriented systems. The other is the application of SOA to M&S, e.g., using service-oriented paradigm to extend the capability of M&S techniques and frameworks. In this work, we pay more attention to the second direction.

### 2.4 Service-Oriented Modeling and Simulation Framework

To define a *service-oriented modeling and simulation framework*, the definition of architecture in software-intensive systems is reviewed [30]: "An architecture is the fundamental organization of a system embodied in its components, their relationships to each other, and to the environment, and the principles guiding its design and evolution." Regarding architecture style, there are two prevailing kinds. One is component oriented, that emphasizes component and capsulation of attributes and functions (e.g., object-oriented systems). The other is connector/relationship oriented, that emphasize the interface, communication protocols and composition between the components (e.g., service-oriented systems).

In the M&S community, Zeigler definite a modeling and simulation framework as [31]: "a framework that defines entities and their relationships that are central to the M&S enterprise. The basic entities of the framework are source system, model, simulator and experimental frame. The basic interrelationships among entities are the modeling and the simulation relationships."

To cover both the software and M&S characteristics, we define a **service-oriented modeling and simulation framework** as:

*A service-oriented modeling and simulation framework is the fundamental organization of a service-oriented simulation system that represents its components (services/simulation services), its components' relationships to each other and to the environment in the system lifecycle of modeling, description, publication, composition, execution, etc., and the principles that guide its design and evolution.*

Because services concentrate more on the capabilities and interfaces than the inner implementation, we address the issues on service-oriented simulation framework while not emphasizing the modeling of services.

**2.5 An Example of Service-oriented Simulation**

Here we take a supposed war game simulation as an example to show the related concepts and processes of service-oriented simulation.

(1) A military enterprise, e.g., DoD, intend to utilize M&S technique to evaluate the effectiveness of the missile defense system in net-centric environment. This may provide a reference to the design and deployment of real systems. – All systems must have a requirement.

(2) The war game environment consists of a certain threat (i.e., an attacking missile), the observation services (i.e., satellite and radar), orient and decision service (i.e., Command and Control system, C2), and defense service (i.e., intercepting missile). DoD want to use a certain simulation runtime infrastructure to connect all the model services and execute them. DoD want to reuse and compose legacy military service/model/simulator while not to develop a new system from the scratch. DoD want to integrate heterogeneous, distributed components. – Requirement analysis, scenario or experimental frame development, identify conceptual services.

(3) DoD search the service broker for required model services. DoD search ontology library for exact terms of Quality of Service (QoS) parameters (e.g., the azimuth angle of radar). The service broker supports vague search by Web browser. – User search the broker for required services

(4) A list of candidate services is acquired. DoD select the exact matched services by other QoS information (e.g., price, reliability, access scheme). DoD get the service description of some completely matched services (e.g., missile, satellite, and C2). – Get service description with URL.

(5) The candidate radar services are not identically matched (e.g., the detect range of radar is not satisfactory.) DoD did not find any defense missile services. – No required service.

(6) DoD write the ads of required services for a radar and a defense missile. – Required service specification.

(7) DoD put the ads to the service broker or other directories.

(8) One radar manufacturer and one defense missile manufacturer find the ads. They believe they can provide the required services. – Service provider.

(9) The radar manufacturer improves the legacy radar service and provides the required service description to the broker. The defense missile manufacturer produces and submits its service description to the broker. – The provider register their service to broker.

(10) The broker informs DoD. DoD have found all required model services. – Get all the implementation-related model service.

(11) DoD search for simulation infrastructure services. They adopt the HLA Evolved Web Service RTI. - Get the implementation-related simulation service.

(12) DoD compose and integrate all the services in a workflow manner. – Service orchestration or static composition.

(13) DoD confirm the availability of all the services. The services authenticate the user (i.e., DoD).

(14) Simulation executes over Web. Messaging through HTTP, SOAP etc.

(15) DoD get the result. It shows the range of the defense missile impacted on the defense effectiveness.

(16) DoD search for defense missile services with a higher range.

(17) DoD find required service. Meanwhile, DoD find a better radar service with higher reliability and lower price. – QoS management.

(18) DoD replace the old defense missile and radar services before execution or during real-time execution by some service agents. – The replacement, chorography or dynamic composition of services.

(19) DoD get required results. The selection and deployment of proper equipments can be recommended according to the simulation results.

(20) DoD pay fees for delivered services.

(21) DoD save the services contact/description for further use.

This is a typical example of military engagement. In particular, the Observe-Orient-Decide-Act (OODA) [32] model depicted by this example is widely adopted both in military domains and others beyond. Even for the war of the old Rome, the threat, observation, orient/decision, and act services also effect (e.g., enemy, human vision, leaders, and arrows). The differences lay on the parameters or QoS. Services live much longer than their implementation.

## 3. Three-dimensional Reference Model for Service-Oriented Simulation

### 3.1 Overview of the reference model

Similar to the 3D morphology of systems engineering [33], research on service-oriented simulation involves (at least) three distinct, yet related fundamental dimensions (domains or viewpoints): M&S, service-orientation, and software/systems engineering. These three dimensions comprise a reference model of service-oriented simulation (Figure 1). The model represents the state-of-the-art and can be applied as an engineering reference model or a framework for the analysis, design, implementation, evaluation, and evolution of service-oriented simulation systems.

As stated before, there are two directions in the service-oriented simulation domain [14]. One is the use of SOA in M&S, i.e., employing a service-oriented paradigm to extend the capacity of M&S techniques and frameworks. An example is the design and implementation of simulators that are services themselves, and can be invoked via SOA protocols [34]. The other is a vice versa approach, where M&S is used for SOA, i.e., M&S techniques are applied to address the problems in service-oriented systems. An example is the application of simulators that evaluate models of software packages designed along the SOA paradigm [35-36]. Similarly, there are two levels in service-oriented simulation: the problem to be simulated, and the simulation mechanism. Both can be service oriented. The reference model is intended to cover both directions and both levels by using different results that are Cartesian products from different orders of M&S and service-orientation dimensions.

Regarding the number and layout of dimensions, there may exist multi-dimensions, sub-dimensions or negative dimensions [37]. Service-oriented simulation must cover at least three dimensions such as M&S, service orientation, and engineering, as explained in the following subsections. In addition, other dimensions or sub-dimensions such as systems of systems [38] and different levels of interoperability [13] exist. However, it is hard to imagine and understand issues generated beyond 3D. Thus, for simplicity, we do not include them here. Furthermore, any additional elements can be regarded as parts of the main three dimensions (e.g., systems of systems and levels of interoperability can be complementary to the engineering dimension). Moreover, the service orientation dimension is broken down along a positive and negative axis. Hence, we stick to the three dimensions depicted in Figure 1.

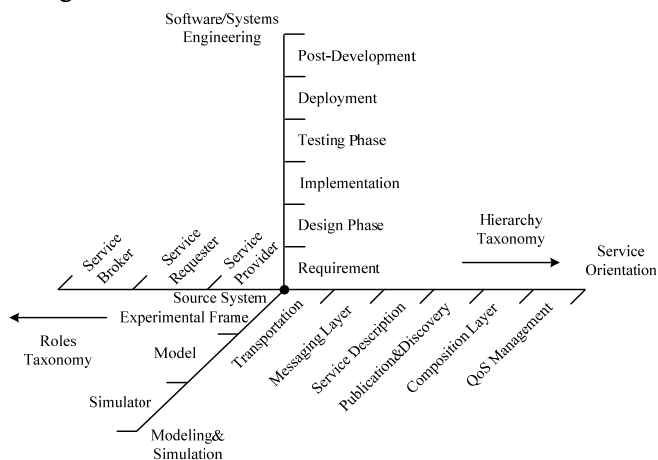

**Figure 1. A Reference Model for Service-Oriented Simulation**

### 3.2 One-Dimensional Implication

A 1D view enables us to look at each fundamental dimension individually. The source system is located at the origin. It stands for the existing or proposed system that we intend to observe or test.

#### 3.2.1 M&S dimension

Besides the source system, the basic entities in M&S [31] include the model, simulator, and experimental frame (EF). Modeling and simulation are the fundamental relationships. Following the logical flow of general M&S practice, EF is firstly determined as the operational formulation of the M&S objectives. Then, we develop a model to represent the source system under a certain EF. Finally, we use a simulator to execute the model to generate its behavior for further study. Hence, we use the sequence of EF, model, and simulator in this dimension. Major research activities in M&S dimension include:

- to represent and model a physical, mathematical, or logical representation of a system, entity, phenomenon, or process correctly;
- to execute models correctly and efficiently;
- to capture the conditions under which the system is observed or experimented;
- to perform experimental design and scenario generation; and
- to collect and validate the outcome of experiments.

Related techniques are listed, such as:
- various modeling approaches,
- various simulation execution approaches and algorithms in the local, parallel or distributed execution paradigms,
- experimental design and scenario generation,
- data collection methods, and verification, validation, and accreditation (VV&A) methods for models and simulators.

#### 3.2.2 Service-orientation dimension

Service-orientation has been an increasingly state-of-the-art and promising approach to design simulation systems. With the appealing characteristics of reusability and interoperability etc., services have been successful in systems analysis, design, development, and integration [4]. The implementation-independent services description can be published by a service provider in the service broker. Based on the published information, a service requestor can discover and compose requested services with other services. Service-oriented approaches can benefit business systems and others in addressing the requirements of agility and flexibility while allowing for changes in the requirements themselves. The SOA [4] is a conceptual framework for the design of business enterprise systems while Web services [39] is the prevailing technology to implement SOA. Previous work [4] provides a detailed review of approaches, technologies, and research issues in service-oriented approaches.

Service-orientation dimension has two taxonomies that come from the conceptual structure of SOA and the implementation hierarchies of Web services, respectively.

The two taxonomies are complementary and the combination of them can better facilitate the analysis and implementation of service-oriented applications. One of the taxonomies, from the viewpoint of roles, is structured as a triangle that consists of a service provider, requester, and broker. We use this particular order for this scale because the service provider and requester are more fundamental roles than the service broker. The service provider must provide its service earlier than the requestor's demand so as to compose a successful application.

The other taxonomy, from the perspective of Web service stack, is where the hierarchies of transportation, messaging, service description, service publication and discovery, composition and collaboration, and quality of service (QoS) management appear. Transportation, messaging and service description are the core layers that constitute the basis for static SOA. Service publication and discovery, composition and collaboration levels enhance the dynamic capabilities for dynamic SOA. QoS management makes services more dependable and robust by focusing on QoS requirements [40] such as performance, reliability, scalability, interoperability, and security. We sequence the elements by their decreasing importance on the scale in Figure 1.

### 3.2.3 Software/systems engineering dimension

Simulation systems usually include software, at least in part [41]. The "Simulation as software engineering" mode of simulation practice [42-43] is applicable for teams of modelers and researchers, lengthy lifecycle and complex projects. For example, this mode dominates the military simulation because of large scale models, long period development and expectation to be reused over a long period. The research and techniques on software engineering, especially software architecture and lifecycle, are of great help to simulation systems. The investigation of McKenzie et al. [41] shows that there are no fundamental difference at the architecture level between simulation systems and general software systems. Formal and informal software architecture design methods can also be widely used in the M&S community.

Additionally, systems engineering can also benefit service-oriented simulation as valuable complement [44] in the hardware, optimization, trade-off, decision making etc. aspects that beyond the scope of software engineering.

The lifecycle of software/systems engineering may be assigned to different ontologies from multiple viewpoints [30,45]. In this work, we use the taxonomy of requirement, design (e.g., description, design, analysis, etc.), implementation, testing, deployment, and post-development (e.g., maintenance, evolution, reuse, etc.). In fact, the activities along the engineering dimension are often cyclic or concurrent.

The research and practice of software/systems engineering are reported in [38,46-47]. Note that design and implementation often receive preferential treatment in the general research and practice.

### 3.3 Two-Dimensional Implication

While 1D looks at each dimension individually, a 2D view inspects each domain constituted from the Cartesian products of every two dimensions. It reveals the systematically cross-discipline landscape of service-oriented simulation. It can also reveal the gaps in the current state of service-oriented simulation systems design.

For a given specific framework that is compatible with the reference model, the issues that result from the reference model are identified as the following three categories:

**(1) core** issues (C), the fundamental nature of service-oriented simulation, if they are not present, the framework cannot be called a service oriented simulation framework;

**(2) supporting** issues (S), the important characteristics of service-oriented simulation, if they are missing, the framework will be heavily impacted; and

**(3) nice-to-have** issues (N), the complementary functions of service-oriented simulation, if they do not appear, the framework may be slightly impacted.

Furthermore, the crossover between research disciplines can be identified and analyzed in Tables 1, 2, and 3. The classification of issues can be applied to both 2D and 3D views.

### 3.3.1 Narrow service-oriented simulation

The Cartesian product of M&S and service orientation dimensions lets us treat the service oriented simulation in a narrow sense. It has two implications that reveal the two directions of SOA for M&S and vice versa, respectively: an approach that enables an extension of the traditional M&S artifacts with the service-orientation principles, and an approach that models or simulates service-oriented systems by means of M&S. Similarly, as for the previous discussion, the Cartesian product of different dimensions provides different directions. It is the fundamental domain of service-oriented simulation. We call it a 'narrow approach' because it lacks rigorous engineering principles or processes. Some ad-hoc research or practices [48-50] belong to this category.

### 3.3.2 M&S engineering

The Cartesian product of M&S and software/systems dimensions provides an M&S engineering domain. It applies engineering principles and methods to traditional M&S as in, for example, the classical HLA Federation Development and Execution Process [51] and VV&A standards [52]. It is the traditional M&S engineering domain that does not necessarily refer to service-oriented simulation.

### 3.3.3 Service-oriented engineering

The Cartesian product of service-orientation and software/systems dimensions creates a service oriented engineering domain. Here, engineering principles are applied to a service-orientation community. Although the basic engineering principles seem still unchanged (along the classical engineering dimension), new requirements and challenges are introduced by the SOA paradigm. For example, services are key elements, service interfaces, reuse and composition are paid more attention to, and the development style is mainly model driven. Service oriented engineering is a new emerging domain. Typical examples include service-oriented systems engineering [10] and service-oriented software engineering [11-12]. In particular, these authors discussed the impact of the SOA paradigm on classical software/systems engineering principles and practices.

**Table 1 Narrow Service-oriented Simulation (M&S vs Services)**

|    | Broker | Requester | Provider | Transport | Messaging | Description | Publish&Discovery | Composition | QoS |
|----|--------|-----------|----------|-----------|-----------|-------------|-------------------|-------------|-----|
| EF | N | N | N | N | N | N | N | N | N |
| M  | S | C | C | C | C | C | S | S | N |
| S  | S | C | C | C | C | C | S | S | N |

The increasing gray intensity of the cells identifies nice-to-have (N), supporting (S), and core issues (C), respectively. EF= Experimental Frame

**Table 2 M&S Engineering (M&S vs Engineering)**

|    | Requirements | Design | Implementation | Testing | Deployment | Post-development |
|----|--------------|--------|----------------|---------|------------|------------------|
| EF | N | N | N | N | N | N |
| M  | S | C | C | S | S | N |
| S  | S | C | C | S | S | N |

The increasing gray intensity of the cells identifies nice-to-have (N), supporting (S), and core issues (C), respectively. EF= Experimental Frame

**Table 3 Service-oriented Engineering (Services vs Engineering)**

|   | Broker | Requester | Provider | Transport | Messaging | Description | Publish&Discovery | Composition | QoS |
|---|--------|-----------|----------|-----------|-----------|-------------|-------------------|-------------|-----|
| Requirements | S | S | S | S | S | S | S | S | N |
| Design | S | C | C | C | C | C | S | S | N |
| Implementation | S | C | C | C | C | C | S | S | N |
| Testing | S | S | S | S | S | S | S | S | N |
| Deployment | S | S | S | S | S | S | S | S | N |
| Post-development | S | S | S | S | S | S | S | S | N |

The increasing gray intensity of the cells identifies nice-to-have (N), supporting (S), and core issues (C), respectively.

### 3.4 Three-Dimensional Implication

Despite of the partial perspective from the 1D and 2D interpretation, a 3D view illustrated in Figure 2 provides a complete multi-perspective consideration of a service-oriented simulation. The whole 3D space is constituted by the Cartesian product of all the dimensions. This represents 'service-oriented M&S engineering', also called 'general service-oriented simulation' because it applies engineering principles to the whole development lifecycle of service-oriented simulation systems. To evolve as a new and mature M&S paradigm, service-oriented simulation must cover the whole 3D space demanded by the 3D model. The importance of the respective cells in the 3D Cartesian product space is identified according to core, supporting, and nice-to-have classification.

The 3D conceptual model can be applied to separate concerns and used as a taxonomy of the existing service-oriented simulation frameworks. Moreover, it can aid domain experts to define clearer and more specific activities. It can also help discover potential new research issues for multiple discipline experts so that sub-phases or steps can then be added using the Cartesian products. Examples of possible research problems generated by crossing the service-orientation and M&S dimensions include how to capsulate the capability of models, simulators, and experimental frames as services, and how to manage, use, and implement them at respective layers. From the engineering point of view, the properties, design, and implementation problems should be considered as complements to the above issues.

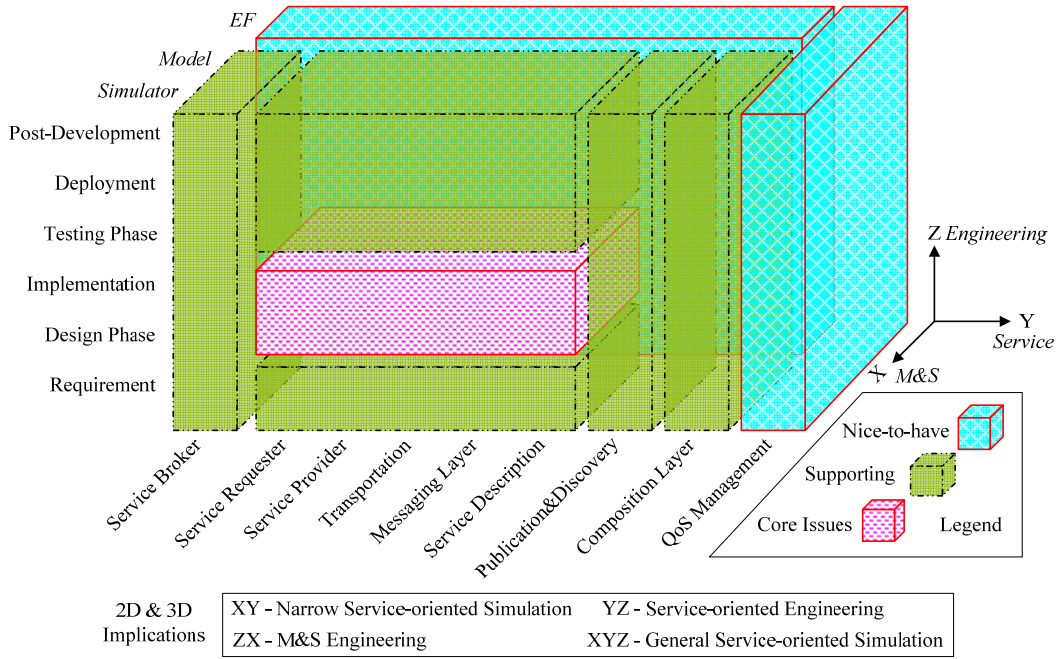

**Figure 2. Three-Dimensional Reference Model for Service-oriented Simulation**

## 3.5 Descriptive and Prescriptive Roles

Engineering methods distinguish characterization (description) and mandatory (prescription) [53]. The 3D reference model for service-oriented simulation can also serve both functions. The descriptive role of the 3D model can be used to describe the ability or maturity of service-oriented simulation frameworks e.g. those surveyed in the following section. It can show the coverage of issues addressed in the 3D space by ad-hoc service-oriented simulation frameworks. The prescriptive role can be utilized to prescribe the issues or requirements that can be satisfied to cover the full 3D space. It can show the values of gaps or strategic future directions of each approach.

The two roles can be applied to show the potential and possible future directions of the classical service-oriented simulation frameworks. It emphasizes applying rigorous engineering principles and methods to embrace the full potential of service-oriented simulation.

## 4. Several Classical Service-oriented Simulation Frameworks

Based on related concepts and three-dimensional reference model of service-oriented simulation, this section describes the state-of-the-art of several classical service-oriented simulation frameworks. They cover some aspects of the whole space depicted with the three dimensional model. They respectively have advantages and limitations as well.

### 4.1 Formalism-based framework

This kind of framework depends on certain simulation formalism in a theoretic or mathematical way. Discrete Event System Specification (DEVS) is a typical example including the following progress.

### 4.1.1 DEVS Unified Process framework (DUNIP)

DEVS Unified Process framework (DUNIP) [15] was proposed by Mittal for integrated development and testing of service-oriented architectures.

**From the M&S perspective**, using DEVS as a unified model specification, they investigate the automated generation of DEVS models from a number of different formalisms such as state-based, rule-based, BPMN/BPEL-based and DoDAF-based. **From the simulation point of view**, DEVS/SOA [34] depicted in Figure 3 was proposed as a simulation service platform to address simulator compatibility issues like DEVS/C++, DEVSJAVA, DEVS/RMI etc. The simulation processes are totally transparent to model execution over the net-centric infrastructure. Users can execute models over Internet by Web services and SOA protocols. The composition and execution of models conforms to System Entity Structure (SES), modular, hierarchical DEVS specification, and DEVS simulation protocols [31].

**From the service-orientation point of view**, DEVS models are regarded as resources while simulators as Web services. DEVS Modeling Language (DEVSML) [54] was proposed to present DEVS models with XML format. The hierarchical architecture of DEVSML is reported in [54]. An approach of abstract wrapper which automatically generates the DEVS Web Service from WSDL interface is presented in [55]. The abstraction mechanism of a coupled model as an atomic model with DEVS state machine and the implementation - the adapter Digraph2Atomic are reported in [15]. A coupled model can be executed like an atomic model. Hence, simulator services are enough to execute DEVS models over net-centric environment without coordinator services. The early version of DEVS/SOA uses centralized communication mechanism by central coordinator. The latest version utilizes direct and real-time communication among services [56].

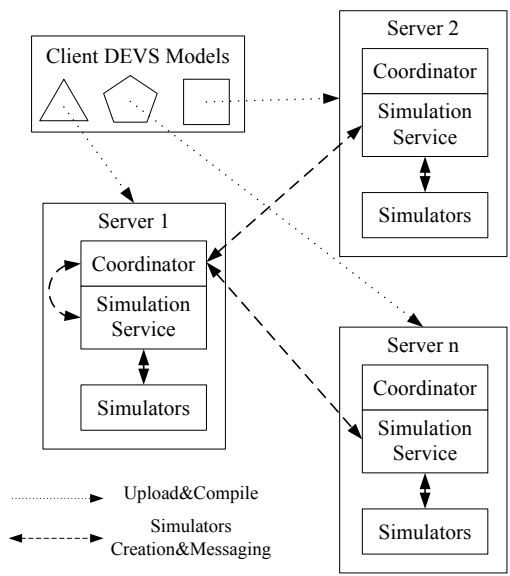

Figure 3. DEVS/SOA architecture

**From the viewpoint of software/systems engineering**, the lifecycle of bifurcated model-continuity methodology was proposed in DUNIP to unify the concepts of model-continuity and M&S framework. The complete process of DUNIP starts from the automated generation of DEVS models from various requirement specifications. Then, DEVS models are transformed to platform-independent XML format using DEVSML. The DEVS/SOA simulation platform is used to deploy, simulate DEVS models and collect output. The architecture and processes of DUNIP are shown in [57].

DUNIP has been partly applied in several projects [15] e.g. Joint Close Air Support (JCAS) Model, DoDAF-based Activity Scenario, Link-16 ATC-Gen Project at JITC, GENETSCOPE Project at JITC. DEVS/SOA and DUNIP are important infrastructures for the net-centric information exchange and systems of systems interoperation [58-59].

### 4.1.2 DEVS simulation framework for service-oriented computing systems (SOAD)

Because of the missing support for some basic SOA concepts in most M&S frameworks, there exist difficulties when modeling and simulating service-oriented computing systems. Hence, Sarjoughian et al. propose an SOA-compliant DEVS (SOAD) simulation framework [60] to address these issues. DUNIP Web enables DEVS framework as service-oriented frameworks but the M&S objectives are not necessary service-oriented systems. While SOAD may not be service-orientation itself, however, the M&S objectives are service-oriented systems. The conceptual framework of SOAD is reported in [60].

**From the service-orientation's perspective**, the research in SOAD concerns the three roles in SOA, messaging patterns, primitive and composite service composition, and hardware model for router link. **From the viewpoint of M&S**, the comparison and contrast between SOA and DEVS are performed. DEVS framework is extended to support the concepts and capabilities of SOA. The basic SOA roles, the modeling of primitive and composite service composition are investigated. Then, the hardware model of network is introduced as valuable complement to the software aspect of SOA. Finally, SOAD is implemented in DEVSJAVA environment and an example is illustrated to show the feasibility. Ramaswamy [61] and Kim [62] model the roles and messages in SOA with classical DEVS formalism. The simulation experiments of a publish/subscribe SOA system are conducted to show the effectiveness. The **software/systems engineering** issues are not the focus of SOAD.

### 4.1.3 Web services based Cell-DEVS framework (D-CD++)

Wainer et al. [16,63] investigates the Web services based Cell-DEVS framework. Cell-DEVS is a DEVS-based formalism that defines spatial models as cell spaces. Web enabling CD++, which is a M&S toolkit to execute Cell-DEVS models, can expose simulation functionalities as Web services to improve interoperability and reusability for users' convenience. The architecture of Web services based distributed simulation framework D-CD++ is shown in [16]. The set of service interfaces in D-CD++ includes session management, configuration, simulation modeling and control, and retrieving data interfaces. The execution of D-CD++ conforms to parallel DEVS simulation protocols and adopts global conservative time management strategy. The master and slave coordinators are used to reduce the number of exchanged messages among simulation services. The experiments and performance analysis are performed for D-CD++ both over the Internet and dedicated fiber optic link. It shows that the overhead of SOAP messaging is the major bottleneck.

**4.1.4 Other related work and a summary**

Other related work include the non-hierarchical DEVSCluster-WS [64] based on Web services and variable structure DEVS [65] as the basis for dynamic SOA. The practice of SOA-based DEVS involve the testing of I/O behaviors in services or systems [66], and network behavior analysis [67-68]. Sun [69] improved DEVS/SOA framework and investigated state management, time management, and messaging scheme. However, the effectiveness, performance, and application of the framework need to be improved.

To summarize, formalism-based (e.g., DEVS) service-oriented simulation framework has the **advantages** of rigorous theory basis and mathematical semantics. It is a general and flexible formalism framework that can model and simulate various systems. Many other formalisms and techniques (e.g., Petri net, state machine, UML, DoDAF) can be transformed or mapped into DEVS formalism [15,70-71]. The specification of DEVS and DEVSML for models, DEVS simulation protocols with interface specification between simulators and models, the system entity structure, dynamic DEVS and research on SOA provides solid foundation for service-oriented simulation. However, DEVS framework has the possible **limitations** of too abstract and hard formalism to follow by users. There also exist difficulties to interoperate with models and simulators in other formalisms. Although DEVS standard organization is trying to standardize DEVS formalisms, model representation, model-solver interface, and model libraries [72-74], they have not been mature, widely recognized and used as standards by industry and academia yet. DEVS primary focus is the education purpose. Hence, the human computer interface, simplicity, convenience, and performance need to be improved.

**4.2 Model-driven framework**

Framework of this type utilizes high level abstract models as the start and basis for the analysis, design, implementation, deployment, and maintenance in the whole lifecycle of service-oriented software development. Dynamic Distributed Service-Oriented Simulation Framework (DDSOS) [17,35-36] is the typical example. DDSOS is a distributed multi-agent service-oriented framework based on Process Specification and Modeling Language for Services (PSML-S) [75]. It has the distinct functionalities such as dynamic simulation federation configuration management, automated simulation code generation, automated code deployment, multi-agent simulation for reconfiguration and dynamic analysis. It is an M&S framework that supports rapid simulation, development, and evaluation of large scale systems. Jia et al. [48] proposes a similar framework. However, the differences between Jia's framework and DDSOS are the replacement of PSML with UML as the common model specification and also the lack of some dynamic properties.

**From the M&S point of view**, PSML-S is taken as the modeling language for SOA systems. The mappings from SOA and SOA workflows to PSML elements, structure models, and PSML models are reported in [76]. The mappings from HLA federation rules and interface specification to PSML are also investigated. RTI is taken as the runtime infrastructure. The optimistic time synchronization approach is used in the simulation engine with the consideration of deadlock, synchronization, dynamic re-composition, and reliability.

**From the service-orientation's perspective**, the services in DDSOS include system simulation agent services, environment simulation agent services and RTI services. Once an application has been developed and deployed by DDSOS, three levels of reconfiguration are available which are rebinding, re-composition and re-architecture. The dynamic properties of DDSOS are achieved by the core ideas of Model Driven Architecture (MDA). The simulation code can be automatically generated, deployed, and executed by the modification of PSML-S models.

**From the viewpoint of software/systems engineering**, the whole lifecycle is supported including modeling and specification, verification, code generation, validation, assembling and deployment, execution and monitoring, evaluation, reconfiguration. The architecture and processes of DDSOS are reported in [17]. DDSOS can completely support service-oriented systems engineering [10]. From the application perspective, according to our knowledge, DDSOS has only applied to some preliminary cases [17,76-77].

The idea and **advantages** of model-driven, excellent dynamic composability and the full support of service-oriented systems engineering build up the solid foundation for service-oriented simulation. The **constrains** exist that the workflow-based behavior models of PSML [78] are incapable of representing simulation systems that are not based on processes. In addition, PSML is not a widely recognized standard. DDSOS focuses more on the domain of service-oriented software development. It only extends some functionalities of RTI in simulation community. It also lacks some high level formalism or theory basis for PSML and DDSOS framework. DDSOS provides dynamic properties. Meanwhile, it also brings difficulties to the efficiency, cost, and implementation. There is no support of mappings and automated transformation from other formalisms, for example from UML to PSML. The practice of DDSOS also needs to be extended.

### 4.3 Interoperability protocol based framework

This approach is based on the some interoperability protocol (e.g., HLA) [51,79-82] as the simulation bus for service integration and information exchange. The typical example is service-oriented HLA (SOHLA) [13]. *Service-oriented HLA refers to the architecture enabled by SOA and Web Services etc. techniques which supports distributed interoperating services.* According to the layers of HLA, Web enabling HLA can be implemented at four layers: at the communication layer (such as Web-Enabled RTI [83-84]), at the interface specification layer (e.g., HLA Evolved Web Service API [85] and Unified Architecture [86]), at the federate interface layer (such as HLA Connector [86]) and at the application layer (e.g., HLA Island [85]). In the Swedish "HLA and SOA integration" in support for the network-based defense, the prototypical architecture has been implemented and tested. This allows service-based and HLA-based systems to interoperate as shown in Figure 4 [86]. This project integrates four federates using the native API, WS API and HLA Connector respectively, which shows the feasibility of those approaches. At present, HLA Evolved Web Service API is the latest progress using SOA and Web services technique to extend the HLA at the interface specification. The new generation HLA standard [87-89] is under ballot by IEEE and will be accepted in recent years. Many leading commercial RTI corporations including Pitch and MAK are playing an active role in revising new HLA standard and developing or has released new versions of RTI [90-92]. Wang et al. [13] surveys the latest research and practice of service-oriented HLA.

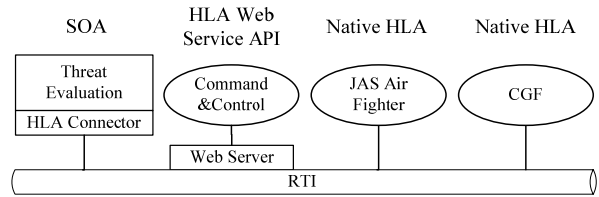

**Figure 4. Architecture of Swedish "HLA&SOA integration" in support for network-based defense**

**From the M&S's perspective**, Base Object Model (BOM) [93], modular Federation Object Model (FOM) [88,94-95] and other enhancements improve the composability and flexibility of HLA simulation systems. **From the viewpoint of service-orientation and simulation**, service-oriented HLA reflects the idea of "simulation as services" [96]. New improvements such as HLA Evolved XML Schema [97], smart update rate [98] and fault tolerant mechanism [99] provide techniques to deal with problems of service-oriented HLA in net-centric environment. **From the perspective of software/systems engineering**, the Federation Development and Execution Process (FEDEP) needs to be modified to reflect the idea of Web centric [100] and support of reuse, composition, and collaboration of services.

Service-oriented HLA has the **advantages** of a set of world wide recognized IEEE standards and also widely used tools and applications. Many future or legacy HLA-compliant simulation resources can be easily modified and reused in the new HLA standard. HLA has solid research and practice foundation both in the academia and defense community. The recent peer survey [101] also reveals that the practical relevance and revision of HLA(e.g. HLA Evolved [87]) are still the future trends in distributed simulation. The **limitations** of service-oriented HLA includes that HLA Evolved is the revision while not the revolution of HLA. The principles and semantics of HLA have not exchanged. Some fundamental rules (e.g. monolithic FOM at the syntactic level) may constrain the further development of HLA. In addition, HLA only focuses on simulation interoperability while not the composability of models or services. It also lacks of rigorous theory foundation. There also exist conflicts between coarse-grained services in SOA and fine-grained services in HLA. Additionally, HLA has the poor capability to support the composability and interoperability at semantic, pragmatic, dynamic, and conceptual levels [102]. Some additional disadvantages and possible future directions are reported in [13].

### 4.4 EXtensible Modeling and Simulation Framework

XMSF [18,103] is defined as a composable set of standards, profiles, and recommended practices for Web-based M&S. XMSF utilizes Web services and related techniques to build up a common M&S technique framework. With the openness, dynamics, maturity, scalability of Web services etc. techniques, the M&S can be integrated with operational systems in the GIG environment. Web/XML, Internet/Networking and M&S are regarded as the major focus areas of XMSF. They cover the M&S and service-orientation dimensions and have their requirements, focus, and related standards respectively [103-105]. SISO also builds up XMSF study group to deal with these issues.

The practice of XMSF includes the Web-Enabled RTI [106] and the project using XMSF to connect Navy Simulation System, Simkit, and CombatXXI for joint modeling and analysis sponsored by SAIC [107]. The Armed Forces of Korea also investigates the intelligent-XMSF approach base on autonomous Web Services [108].

The common technique framework of XMSF provides conceptual and technical support for service-oriented simulation. The related profiles of XMSF also provide experience in practice and implementation. The **limitation** of XMSF is its lack of concrete standards and implementation for service description, composition, and integration. It also lacks the support of software/systems engineering. In addition, the XMSF project has been terminated due to the lack of financial support for XMSF study group in 2005. The product development group was not built up to develop the standards and products for XMSF [109]. The focus of the members in XMSF changes to the research on GIG and HLA Evolved in net-centric environment.

**4.5 Open Grid services architecture based framework**

The Grid is to integrate various distributed resources as "Grid" to support the sharing of collaborative resources and problems solution for virtual organizations. Resources sharing is the essence of the Grid [110]. Grids can be classified into computing, storage, data, knowledge, and service etc. Grids according to the properties of resources on the nodes [111]. There are two connections between SOA and Grid computing. One is the application of SOA in Grid computing, that is, the Grid architecture such as the leading OGSA is based on SOA. Another is the application of Grid computing in SOA, that is, services are taken as Grid resources by Grid computing techniques to form service Grid, which supports sharing, management and convenient access to services.

**From the service-orientation perspective**, Web services focus on the interface description and messaging of services. While Grid computing emphasizes the distributed computing resources including transparent access, fault tolerance, load balance etc. The two techniques are complementary and move towards their unification. OGSA is the architecture based on Web services and techniques. Grid service is the extension of Web services, which support the stateful services. The new specification called the Web Services Resources Framework (WSRF) [112] takes classical Web services as the interface to the stateful resources. **From the M&S point of view,** a framework called SOAr-DSGrid is proposed by Nanyang Technological University Singapore for developing a component-based distributed simulation and executing the simulation in SOA on the Grid [113]. The Grid-based HLA management system (G-HLAM) by Rycerz et al. [114-115], the service-oriented HLA RTI framework called SOHR by Ke Pan et al. [116], and the research of Xie et al. [117] implement HLA/RTI services as Grid services. Other computing and storage resources concerning to M&S can be implemented as Grid services too. Zhang et al. [118] gives a detail survey on the Grid-based distributed simulation. Zhang's dissertation [119] investigates the HLA RTI service, resource discovery service, simulation execution service, and simulation task migration issues in OGSA environment. Li et al. proposes a service-oriented GRID simulation called Cosim-Grid [19,120]. It is a service-oriented simulation framework based on HLA, Product Lifecycle Management (PLM) and Grid/Web services. It improves HLA on dynamical share, autonomy, fault tolerance, capability of collaboration, and security mechanism. The prototype Cosim-Grid includes resource layer, Grid resource service middleware layer, simulation application oriented middleware layer, application portal layer of simulation Grid and application layer. Cosim-Grid extends the practice of OGSA based simulation framework. Li et al. summarizes the essence, architecture, key techniques and practices of simulation Grid in their latest review [121].

OGSA is a **valuable** complement to the state-of-the-art of service-oriented simulation framework from the resource management point of view. The research on Grid simulation provides foundation for the reuse, distribution and management of model components and other resources. The advantages also include the dynamic allocation and fault tolerance of resources, and also the transparency of computing resources to the users. **However**, the M&S theory foundation, software/systems engineering, performance, and making full use of SOA in OGSA based service-oriented simulation frameworks need further research. Moreover, OGSA needs some middleware such as Globus, while Web services use open commercial standard and technique. Additionally, the reliable and easy to use, and persuading numerous institutions to open their resources to the outsiders are to be improved as well.

**4.6 Other service-oriented simulation framework**

Besides the above classical service-oriented simulation frameworks, Northrop Grumman's Service Integration/Interoperation Infrastructure (Si3) [49-50] is proposed to support simulation-based transformation. **From the viewpoint of M&S**, the composite simulation applications can be created through the integration and interoperation of models, simulation, applications, tools, utilities, and databases. Si3 also provides a toolset to package applications as self-describing, discoverable, composable, and configurable services. Hence it enables the integration and interoperation of independent, distributed heterogeneous applications. The conceptual and implementation architectures of Si3 are shown in [49]. However there seem few publications known to the authors. The design and details of Si3 need further exploration.

Besides Si3, another service-oriented simulation framework for military use is Web enabled Joint Theater Level Simulation (JTLS) [122]. It is used for conducting large-scale multinational exercises. Simulation operators (JTLS users) can participate in joint trainings through Web browser to communicate with a JTLS game at remote site anywhere in the world. Web enabled JTLS uses centralized model execution and computing style while expose configuration, message and order management functions to Web services. It is a successful service-oriented war game simulation with lower temporal and spatial resolution, lower update rate and slower time advance rate.

In the enterprise application integration domain, international standard organizations and many researchers investigate the service description, publish and discovery, messaging and quality of services (QoS) based on Web services and semantic Web services [123-127]. The related standards and publications build the foundation of specification and supporting techniques. However, these researches focus on the **service-orientation's perspective**. The models, simulators, time management etc. issues in the M&S community need further research.

The research of Zhang's [27] and Song's [128] belong to the ontology or semantic driven service-oriented simulation framework. Ontology or semantic Web are used to improve the communications between users and Web services that using different terminologies [129]. They propose the conceptual framework base on ontology or semantics, investigate the issues of simulation service description, discovery, matchmaking, QoS-driven simulation services composition, dynamic simulations services composition, and fault-tolerance. However, they pay more attention to the service-orientation dimension (e.g. Web services or semantic Web services). The issues in the M&S dimension, such as the VV&A of simulation services composition, states management and time management of simulation service, the presentation and implementation of model services and simulator services, and the constitution and improvement of ontology library for M&S community deserve further research. In Zeigler's new book [59], ontology and pragmatic framework are introduced to facilitate M&S based data engineering for net-centric environment.

## 5. Comparison of service-oriented frameworks

Based on the 3D reference model, the overall comparison of the classical service-oriented simulation frameworks are illustrated in Table 4. To provide multi-views landscape for stakeholders in different domains, we also give a detailed comparison from 1D, 2D, and 3D views in the appendix Tables A1~A9. The detailed examples applied to DUNIP and DDSOS frameworks are reported in [20].

According to the comparison and survey, we can get some observations. The formalism-based approach has the rigorous theory basis and can support modular hierarchical discrete event systems simulation. However, it has the limitations of standardization and ease to use. The model-driven method pays more attention to service-oriented software engineering and dynamic properties, while has the limited capability in M&S aspect. The interoperability protocol based approach has the mature basis of international standards, widely application and promising potential. However, the model support and the higher levels of interoperability need to be improved. The XMSF outlines the techniques framework of Web-based simulation. However, it needs the concrete implementation, and the research of XMSF study group has ceased. The OGSA-based method supports the dynamic management, reuse, and transparency access of various simulation resources. However, it needs the infrastructure of Grid middleware, and the Grid-based M&S theory and approaches need further research. Si3 and ontology-driven frameworks have the advantages of service integration and semantic interoperability. However, the M&S theory and the VV&A of services need to be improved.

**Table 4 The overall comparison of classical service-oriented simulation methods**

| Methods | Examples | M&S | Service-orientation | System Engineering | Advantages | Limitations |
|---|---|---|---|---|---|---|
| Formalism based | DUNIP, DEVS/SOA, SOAD, D-CD++ | Unified DEVS model specification. DEVSML for platform independent models. SOAD can model & simulate service-based software & hardware systems. DEVS simulation protocol. | Simulators as services, models as resources in DUNIP. No coordinators services. Session management, configuration, simulation modeling & control, and retrieving data service interfaces in D-CD++. | DUNIP has bifurcated model-continuity systems engineering methodology. | Mature formalism with long history. Strong presentation capability to various systems. Rigorous theory basis and mathematical semantics. | Too abstract & hard to follow by users. Have not been widely recognized industrial & academic standards. Focus more on education. Simplicity, convenience & performance need to be improved. |
| Model driven | DDSOS | PSML-S can model SOA systems. RTI as runtime infrastructure. Optimistic time synchronization. | Systems/environment simulation agent services & RTI services. Support dynamic rebinding, re-composition, and re-architecture. | MDA and service-oriented systems engineering (SOSE) support. | Model-driven, excellent dynamic composability and SOSE support | Focus on service oriented software development. Limited simulation capabilities. Theory, efficiency & applications to be improved. |
| Interoperability protocol based | Service oriented HLA, HLA Evolved Web Service API etc. | BOM & modular FOM facilitate interoperability levels of models. Low bandwidth, uncertainty & dynamic properties need considering. | Web-Enabling HLA at communication, HLA interface specification, federate interface & application layers. HLA Evolved XML Schema, smart update rate and fault tolerance mechanisms. | FEDEP needs to be modified to reflect the idea of Web centric and support of reuse, composition, and collaboration of services. | World wide recognized IEEE standards. Solid research and practice foundations. HLA Evolved upcoming standards | Revision while not the revolution of HLA may constrain further development. Lower levels of interoperability. Conflicts between SOA & HLA in service granularity. |
| XMSF | XMSF and profiles | The M&S focus area of XMSF. | The Web/XML, Internet/Networking focus area of XMSF. | N/A | Technique framework; related focus areas, issues & techniques. | Lack of concrete standards, products & systems engineering support. Has terminated. |
| OGSA based | Cosim-Grid, SOAr-DSGrid, G-HLAM, SOHR | Simulation components, HLA/RTI services, computing & storage resources can be Grid services. | Focus on management of distributed computing resources. Base on Grid middleware. | Not clear | Resource dynamic allocation, load balance & fault tolerance. Transparency. | Need Grid middleware. M&S theory basis, systems engineering, performance & full use of SOA to be improved. |
| Other approaches | Si3, ontology/semantic driven framework | HLA/RTI simulation engine in Si3. Service description, semantic service matchmaking, not focusing on simulation execution in ontology approach. | Si3 packaging models, simulation, applications, tools, utilities & databases as services. Ontology method focuses on service UDDI, composition & fault-tolerant. | Have some development and usage procedures. | Integration & interoperation of heterogeneous applications in Si3. UDDI & semantic composition in ontology methods. | Few publications & not mature. Need further investigation in the M&S dimension especially VV&A, states & time management of simulations service. |

From Table 4 and also the viewpoint of using SOA paradigm to extend the capability of M&S frameworks, we conclude that the formalism-based approach is the most mature one from M&S theory's perspective. While the interoperability protocol based method is the most potential one in the practice of service-oriented simulation engineering. The combination of the two approaches, and also with the rigorous engineering methods and principles of the levels of conceptual interoperability model (LCIM) [53,130] can better facilitate the meaningful composition and interoperation of simulation services, and promote the theory and practice of service-oriented M&S.

# 6. Conclusion and future work

As the requirements of simulation interoperability, reusability, and composability in the net-centric environment extend continually, simulation systems are developing towards standardization, introduction of components, hierarchy, networks and services abilities [131].

From the academic viewpoint, service-oriented M&S is the interdisciplinary field of M&S, service-oriented paradigm, and software/systems engineering. It stands for the current focus and future directions of M&S in the prevailing net-centric environment. The three-dimensional reference model can be used to identify the supporting domains, research issues and used as a guideline to the engineering lifecycle of service-oriented simulation. The theory and practice of classical frameworks can facilitate the research, design, implementation, and application of service-oriented simulation frameworks through different perspectives or multi-viewpoints. Service-oriented M&S is one of important approaches to deal with ambiguity, uncertainty, and variability [132] of complex systems in the net-centric environment.

From the viewpoint of practice value, service-oriented M&S frameworks can extend the capabilities of classical frameworks such as the reusability, composability, dynamic, extensibility, interoperability, fault-tolerance, intellectual property, deployment, access, maintenance, agility of various simulation resources. Service-oriented M&S framework provides a new solution to the integration, interoperability, and reusability of heterogeneous resources and applications in simulation, enterprise, legacy artifacts, and many other communities. Furthermore, it can also facilitate dealing with real time and uncertain decision and application integration problems in a highly dynamic and agile sphere such as net-centric war game. Moreover, it provides a new idea of dynamic and rapid composition and simulation on demand. The service-oriented approaches have significant benefits to improve the way information and functional capabilities are shared, collaborated, and integrated. It will promote the transformation of current simulation resources and the development of new applications, and have important research value and wide application potential.

Although the research on service-oriented simulation frameworks has made great progress, there are still many unsolved problems and difficulties which can be regarded as future directions:

**(1) The body of knowledge of service-oriented simulation**

To make service-oriented simulation concept in general sense as a successor of object-oriented simulation [28], the body of knowledge [37], rigorous theory basis and related standards, techniques, and implementations need to be built up. The intension and extension of the concept need to be investigated. Moreover, although the formalism-based and model-driven approaches can cover both the directions of M&S for SOA and vice versa, the body of knowledge may not be complete until the gap of the first direction (M&S for SOA) is also comprehensively investigated. The service modeling approaches, VV&A, experiments etc. should also be addressed.

**(2) The three-dimensional reference model**

The 3D reference model can be regarded as a good beginning to investigate the service-oriented simulation paradigm. However, the model needs much future work to become mature. For example, detailed evaluation criteria should be given for each cell in the 3D space. The formal and mathematical presentation and evaluation criteria of the reference model are also welcome. The consistency of the multi views, the qualitative and quantitative roles of the reference model are also interesting issues. Furthermore, domain experts can help to verify the preliminary evaluation shown in Tables A1~A9. They can also define, modify or add more clearly, specific activities in the three dimensional model. Moreover, interdisciplinary experts can utilize the model to discover more new research issues through the Cartesian products of three or two dimensions.

In addition, integration, interoperation, and composition are important properties of service-oriented simulation. Information needs to be exchanged among services to get meaningful comprehension. Hence the LCIM [53,130] and system of systems [38] dimensions can be added to the three dimensional model as valuable complements. The descriptive and prescriptive roles of LCIM can be used to describe the properties or the requirements from the syntactic, semantic, pragmatic, dynamic, and conceptual levels from the information exchanged point of view.

**(3) M&S dimension**

From the M&S point of view, most research known to the authors expose the functionalities of simulation infrastructure as services, which reflect the idea of "simulation as services" [96]. However, from the modelers' perspective, exposing simulation models as services is also necessary and important for intellectual property and model reuse. DEVS models specification, DEVSML and BOM standard provide solid foundation for the standardization of model services. However, the conflicts between reuse granularity [133] and execution performance need to be solved. Moreover, the standardization of simulation services, the VV&A of standalone services and composition services, performance [13], multicast over wide area network [134], semantically lossless interoperation and composition of services [135], the interoperation with Commercial-Off-The-Shelf (COSTS) simulation packages [136] etc. are also interesting research issues.

**(4) Service-orientation dimension**

From the perspective of service-orientation, making full use of SOA [3] such as supporting simulation components for registering, finding and dynamic integration with UDDI, enhancing the service orchestration and choreography etc. properties of dynamic SOA are worth further researching. In addition, the QoS, fault-tolerant, automated or semi-automated service interoperation and composition [137] deserve further investigation.

**(5) Software/systems engineering dimension**

Service-oriented simulation systems should utilize rigorous engineering principles to facilitate the reuse and composition of complex applications. Standard software/systems engineering approaches should be investigated to guide the analysis, design implementation and application of service-oriented simulation frameworks.

**(6) Research and Practice of Service-oriented simulation frameworks**

The classical service-oriented simulation frameworks surveyed in this paper have their pros and cons. Whether using a unified and standard framework or permitting the diversity with the multi-formalism transformation approaches or tools need to be trade-off. The compatibility, reusability, and composability of various service resources developed in specific framework need further research. The combination of the formalism-based and interoperability protocol based approaches can make full use of the advantages both in theory and practice. Sarjoughian and Zeigler [138] investigated the possible combination of traditional DEVS and HLA. However, their combination in net-centric environment needs further research. In additional, the guideline of LCIM can better facilitate the meaningful composition of simulation services.

To summarize, service-oriented simulation looks very promising for M&S in the information age and net-centric environment. In order to be mature and evolved to the new simulation paradigm, it needs the joint effort of researchers and practitioners from all the communities, in particular academia, industry, and government.

# Acknowledgement

The authors would like to thank Prof. Dr. Pieter J. Mosterman, Dr. Justyna Zander, Prof. Dr. Feng Yang, Dr. Jingjie Li, and Dr. Yonglin Lei for their helpful discussions. The authors would like to thank the work of the "Three-Dimensional Morphology of Systems Engineering" and "the Levels of Conceptual Interoperability Model" team that positively inspired our research. This work was supported by the National Natural Science Foundation of China [grant numbers 60674069, 60574056, 60974073, 60974074].

Some sections of this paper are based on the authors' Science Letters: Wang, W. G., W. P. Wang, J. Zander, and Y. F. Zhu. 2009. Three-dimensional conceptual model for service-oriented simulation. Journal of Zhejiang University SCIENCE A 10(8):1075-1081. We are granted that "detailed research articles can still be published in other journals in the future after Science Letters is published by JZUS (Journal of Zhejiang University SCIENCE A)".
http://www.zju.edu.cn/jzus/

**Wenguang Wang** is a Ph.D. candidate at College of Information Systems and Management, National University of Defense Technology (NUDT), China. He is a member of the SCS, SISO and CASS (Chinese Association for System Simulation). He is also an invited reviewer for the international journal Simulation Modelling Practice and Theory. His research interests include service-oriented simulation, High Level Architecture, DEVS, simulation composability and interoperability etc. Email: wgwangnudt@gmail.com

**Weiping Wang** is a professor at College of Information Systems and Management, National University of Defense Technology (NUDT), China. He is the vice dean of the Graduate School, NUDT. He is the founder and director of Systems Simulation Lab, NUDT. He has over twenty years of experience in systems modeling and simulation community. He is a Ph.D supervisor. His research interests include systems simulation, system of systems engineering, systems of systems simulation, simulation based acquisition, simulation composability and interoperability etc. Email: wangwp@nudt.edu.cn

**Yifan Zhu** is a professor at College of Information Systems and Management, National University of Defense Technology (NUDT), China. He is the deputy director of the Department of Systems Engineering, College of Information Systems and Management, NUDT. He is the cofounder of Systems Simulation Lab, NUDT. He has over twenty years of experience in systems modeling and simulation community. He is a Ph.D supervisor. He was a visiting scholar in College of Engineering, Virginia Tech University, USA from 2007 to 2008. His research interests include systems simulation, virtual prototyping, simulation based acquisition, simulation composability and interoperability etc. He is the principal investigator of the National Natural Science Foundation of China project on "Composable Simulation" under grant number 60574056.

**Qun Li** is a professor at College of Information Systems and Management, National University of Defense Technology (NUDT), China. He is currently the director of Systems Simulation Lab, NUDT. He has nearly twenty years of experience in systems modeling and simulation community. He is a M.Sc. supervisor. His research interests include systems simulation, simulation based acquisition, service-oriented simulation, simulation composability and interoperability etc. He is the principal investigator of the National Natural Science Foundation of China project on "Service-oriented Simulation" under grant number 60674069.

# Appendix

**Table A1 The comparison of frameworks from M&S dimension (1D view)**

| Frameworks | Model | Simulator | Experimental Frame |
|---|---|---|---|
| Formalism based | DEVSML, DEVS, SES | DEVS simulation protocol, DEVS/SOA | DUNIP test models/federations |
| Model driven | PSML | HLA-based | System/environment agent service |
| Interoperability protocol based | BOM, SOM, FOM, modular SOM/FOM | RTI, Web service API | N/A |
| XMSF | N/A | N/A | N/A |
| OGSA based | Model resource | Simulator functionalities and resource | N/A |
| Others | Model service, ontology | Simulator service, ontology | N/A |

**Table A2 The comparison of frameworks from service-orientation dimension (1D view)**

| Frameworks | Broker | Requester | Provider | Transport | Messaging | Description | P&D | Composition | QoS |
|---|---|---|---|---|---|---|---|---|---|
| Formalism based | F2 | F2 | F1, F2, F3 | F1, F2, F3 | F1, F2, F3 | F1, F2, F3 | F2 | F2 | F2, F3 |
| Model driven | M1 | M1 | M1 | M1 | M1 | M1 | M1 | M1 | M1 |
| Interoperability protocol based |  |  | I1 | I1 | I1 | I1 |  |  |  |
| XMSF |  |  | X1 | X1 | X1 | X1 |  |  |  |
| OGSA based |  |  | G1 | G1 | G1 | G1 |  |  | G1 |
| Others | O1, O2 |  | O1, O2 | O1, O2 | O1, O2 | O1, O2 | O1, O2 | O1, O2 | O2 |

F1= DUNIP; F2=SOAD; F3=D-CD++; M1=DDSOS; I1=SOHLA; X1=XMSF; G1=OGSA based framework; O1=Si3; O2=Ontology driven framework; P&D=Publication & Discovery

**Table A3 The comparison of frameworks from engineering dimension (1D view)**

| Frameworks | Requirements | Design | Implementation | Testing | Deployment | Post-development |
|---|---|---|---|---|---|---|
| Formalism based | Y | Y | Y | Y | Y |  |
| Model driven |  | Y | Y | Y | Y | Y |
| Interoperability protocol based |  | Y | Y | Y |  |  |
| XMSF |  | Y | Y |  |  |  |
| OGSA based |  | Y | Y |  |  |  |
| Others |  | Y | Y |  |  |  |

**Table A4 The comparision of frameworks from narrow service-oriented simulation (M&S vs Services 2D view)**

|    | Broker | Requester | Provider | Transport | Messaging | Description | P&D | Composition | QoS |
|----|--------|-----------|----------|-----------|-----------|-------------|-----|-------------|-----|
| EF | M1 | M1 | M1 |  | F1 | F1, F3, M1 |  | F1(SES) |  |
| M  | $F2^T$, M1, $M1^T$, G1, O2 | F1 (User), $F2^T$, M1, $M1^T$ | F1 (compile, transform, validate), $F2^T$, M1, $M1^T$, G1, O1 | $F2^T$ (hardware & software) | $F2^T$, M1, $M1^T$, G1 | F1 (DEVSML), $F2^T$, M1, $M1^T$, G1, O2 | $F2^T$, O2 | F1(SES, static), $F2^T$ (static), M1, $M1^T$, O2 | $F2^T$, M1, G1, O2 |
| S  | G1, O2 | F1 (User), | F1 (DEVS/SOA), F3, M1, I1, X1, G1, O1 | F3, I1, X1 | F1, F3, M1, I1, X1, G1 | F1, F3, M1, I1, X1, G1, O2 | M1, O2 | F1(SES, static), M1, O2 | F3, M1, G1, O2 |

The increasing gray intensity of the cells identifies nice-to-have, supporting, and core issues, respectively. Elements marked with a superscript 'T' (transposition) identify M&S for SOA; normal elements identify SOA for M&S. F1= DUNIP; F2=SOAD; F3=D-CD++; M1=DDSOS; I1=SOHLA; X1=XMSF; G1=OGSA based framework; O1=Si3; O2=Ontology driven framework; EF=Experimental Frame; SES=System Entity Structure; DEVSML=DEVS Modeling Language; P&D=Publication & Discovery

**Table A5 The comparision of frameworks from M&S engineering (M&S vs Engineering 2D view)**

|    | Requirements | Design | Implementation | Testing | Deployment | Post-development |
|----|--------------|--------|----------------|---------|------------|------------------|
| EF | F1 | F1, $F2^T$, F3, M1 | F1, $F2^T$, F3, M1 |  | F1 |  |
| M  | F1, M1(PSML), O2 | F1, $F2^T$, M1(PSML), G1, O1 | F1, $F2^T$, M1(PSML), G1, O1 | F1, M1 | F1, M1 | M1, O2 |
| S  | I1, O2 | F1, F3, M1, I1, G1, O1 | F1, F3, M1, I1, G1, O1 | I1 | F1, I1 | F3 (Performance), G1, O2 |

The increasing gray intensity of the cells identifies nice-to-have, supporting, and core issues, respectively. Elements marked with a superscript 'T' (transposition) identify M&S for SOA; normal elements identify SOA for M&S. F1= DUNIP; F2=SOAD; F3=D-CD++; M1=DDSOS; I1=SOHLA; G1=OGSA based framework; O1=Si3; O2=Ontology driven framework

**Table A6 The comparision of frameworks from service-oriented engineering (Services vs Engineering 2D view)**

|      | Broker | Requester | Provider | Transport | Messaging | Description | P&D | Composition | QoS |
|------|--------|-----------|----------|-----------|-----------|-------------|-----|-------------|-----|
| Req  |        | F1 | F1 | $F2^T$ | F1 | F1 |  | M1 |  |
| Dsn  | $F2^T$, M1, O2 | F1, $F2^T$ | F1, $F2^T$, F3, M1, I1, G1, O1 | F1, $F2^T$, M1, I1, G1 | F1, $F2^T$, F3, M1, I1, G1 | F1, $F2^T$, F3, M1, I1, G1, O1, O2 | $F2^T$, M1, O2 | F1(SES, static),, $F2^T$ (static), M1, O2 | $F2^T$, M1, G1, O2 |
| Imp  | $F2^T$, M1, O2 | F1, $F2^T$ | F1, $F2^T$, F3, M1, I1, G1, O1 | F1, $F2^T$, M1, I1, G1 | F1, $F2^T$, F3, M1, I1, G1 | F1, $F2^T$, F3, M1, I1, G1, O1, O2 | $F2^T$, M1, O2 | F1(SES, static),, $F2^T$ (static), M1, O2 | $F2^T$, M1, G1, O2 |
| Tst  |        | F1, M1, I1 |  |  |  | F1, M1 |  | M1 |  |
| Dply |        | F1 | F1, M1 | F1 | F1 | F1 |  |  |  |
| PstD |        |  | M1 |  |  |  |  |  |  |

The increasing gray intensity of the cells identifies nice-to-have, supporting, and core issues, respectively. Elements marked with a superscript 'T' (transposition) identify M&S for SOA; normal elements identify SOA for M&S. F1= DUNIP; F2=SOAD; F3=D-CD++; M1=DDSOS; I1=SOHLA; G1=OGSA based framework; O1=Si3; O2=Ontology driven framework; Req=Requirement; Dsn=Design; Imp=Implementation; Tst=Test; Dply=Deploy; PstD=Post-development; P&D=Publication & Discovery

Table A7 The comparision of frameworks from model's perspective (3D view)

| | Broker | Requester | Provider | Transport | Messaging | Description | P&D | Composition | QoS |
|---|---|---|---|---|---|---|---|---|---|
| Req | | | F1 | F2$^T$ | F1 | F1, M1 | | | |
| Dsn | F2$^T$, O2 | F1 (User), F2$^T$, M1(PSML$^T$) | F1(compile, transform, validate), F2$^T$, M1(PSML$^T$), G1, O1 | F1, F2$^T$, M1(PSML$^T$) | F1, F2$^T$, M1(PSML$^T$) | F1(DEVSML), F2$^T$, M1(PSML$^T$), O1, O2 | F2$^T$, O2 | F1(SES, static), F2$^T$ (static), M1(PSML$^T$), O2 | F2$^T$, O2 |
| Imp | F2$^T$, O2 | F1 (User), F2$^T$, M1(PSML$^T$) | F1(compile, transform, validate), F2$^T$, M1(PSML$^T$), G1, O1 | F1, F2$^T$, M1(PSML$^T$) | F1, F2$^T$, M1(PSML$^T$) | F1(DEVSML), F2$^T$, M1(PSML$^T$), O1, O2 | F2$^T$, O2 | F1(SES, static), F2$^T$ (static), M1(PSML$^T$), O2 | F2$^T$, O2 |
| Tst | | | F1, M1 | | | F1 | | | M1 |
| Dply | | | F1, M1 | F1, M1 | F1, M1 | F1, M1 | | M1 | M1 |
| PstD | | | M1 | M1 | M1 | M1 | | M1 | M1 |

The increasing gray intensity of the cells identifies nice-to-have, supporting, and core issues, respectively. Elements marked with a superscript 'T' (transposition) identify M&S for SOA; normal elements identify SOA for M&S. F1= DUNIP; F2=SOAD; M1=DDSOS; G1=OGSA based framework; O1=Si3; O2=Ontology driven framework; SES=System Entity Structure; DEVSML=DEVS Modeling Language; Req=Requirement; Dsn=Design; Imp=Implementation; Tst=Test; Dply=Deploy; PstD=Post-development; P&D=Publication & Discovery

Table A8 The comparision of frameworks from simulator's perspective (3D view)

| | Broker | Requester | Provider | Transport | Messaging | Description | P&D | Composition | QoS |
|---|---|---|---|---|---|---|---|---|---|
| Req | | | | | F1 | | | | |
| Dsn | M1, O2 | F1 (User), M1 | F1, F3, M1, X1, I1, G1, O1 | F1, M1, X1, I1, G1 | F1, F3, M1, X1, I1, G1 | F1, F3, M1, X1, I1, G1, O1, O2 | M1, O2 | M1, O2 | M1, O2 |
| Imp | M1, O2 | F1 (User), M1 | F1, F3, M1, X1, I1, G1, O1 | F1, M1, X1, I1, G1 | F1, F3, M1, X1, I1, G1 | F1, F3, M1, X1, I1, G1, O1, O2 | M1, O2 | M1, O2 | F3, M1, O2 |
| Tst | | | I1 | | | | | | |
| Dply | | | F1, M1 | F1, M1 | F1, M1 | F1, M1 | | M1 | M1 |
| PstD | | | M1 | M1 | M1 | M1 | | M1 | M1 |

The increasing gray intensity of the cells identifies nice-to-have, supporting, and core issues, respectively. F1= DUNIP; F3=D-CD++; M1=DDSOS; I1=SOHLA; X1=XMSF; G1=OGSA based framework; O1=Si3; O2=Ontology driven framework; SES=System Entity Structure; DEVSML=DEVS Modeling Language; Req=Requirement; Dsn=Design; Imp=Implementation; Tst=Test; Dply=Deploy; PstD=Post-development; P&D=Publication & Discovery

Table A9 The comparision of frameworks from experimental frame's perspective (3D view)

| | Broker | Requester | Provider | Transport | Messaging | Description | P&D | Composition | QoS |
|---|---|---|---|---|---|---|---|---|---|
| Req | | | | | | F1 | | | |
| Dsn | O2 | | M1 | F1, M1 | F1, M1 | F1, M1 | O2 | M1 | F2$^T$, F3, M1, G1, O2 |
| Imp | O2 | | M1 | F1, M1 | F1, M1 | F1, M1 | O2 | M1 | F2$^T$, F3, M1, G1, O2 |
| Tst | | | | | | | | | |
| Dply | | | | F1 | F1 | F1 | | | |
| PstD | | | | | | | | | |

The gray intensity of the cells identifies nice-to-have issues. Elements marked with a superscript 'T' (transposition) identify M&S for SOA; normal elements identify SOA for M&S. F1= DUNIP; F2=SOAD; F3=D-CD++; M1=DDSOS; I1=SOHLA; X1=XMSF; G1=OGSA based framework; O2=Ontology driven framework; Req=Requirement; Dsn=Design; Imp=Implementation; Tst=Test; Dply=Deploy; PstD=Post-development; P&D=Publication & Discovery